\documentclass[sigplan, screen, nonacm]{acmart}

\setcopyright{rightsretained}
\acmPrice{}
\acmDOI{}
\acmYear{2020}
\copyrightyear{2020}
\acmJournal{PACMPL}
\acmVolume{}
\acmNumber{SYNT}
\acmArticle{}
\acmMonth{7}

\acmConference[SYNT '20]{SYNT '20: 9th Workshop on Synthesis}{July 19, 2020}{Los Angeles, CA, USA}
\acmISBN{}

\usepackage{tikz}
\usepackage{listings}
\usepackage{algorithm2e}
\usepackage{enumitem}

\definecolor{codegreen}{rgb}{0,0.6,0}
\definecolor{codegray}{rgb}{0.5,0.5,0.5}
\definecolor{codepurple}{rgb}{0.58,0,0.82}
\definecolor{backcolour}{rgb}{0.95,0.95,0.92}
\definecolor{darkmagenta}{rgb}{0.55,0,0.55}

\newcommand{\uclid}{{\textsc{Uclid5}}\xspace}
\newcommand{\smtlib}{{\textsc{smt-lib}}\xspace}
\newcommand{\sygusif}{{\textsc{s}y\textsc{g}u\textsc{s-if}}\xspace}
\newcommand{\llamalib}{{\textsc{synth-lib}}\xspace}
\newcommand{\keyword}[1]{\textcolor{darkmagenta}{\texttt{#1}}}

\newcommand{\codelike}[1]{\texttt{#1}}

\lstdefinelanguage{uclid}{
  sensitive = true,
  keywords={module, forall, exists, Lambda, if, else, assert, assume, havoc,
            for, range, skip, case, esac, init, next, control, function, procedure,
            returns, call, define, type, var, input, output, const, property,
            invariant, synthesis, grammar, requires, ensures, modifies, instance, axiom, 
            enum, record, integer, boolean, true, false},
  numbers=left,
  numberstyle=\footnotesize,
  stepnumber=1,
  numbersep=8pt,
  showstringspaces=false,
  breaklines=true,
  frame=top,
  comment=[l]{//},
  morecomment=[s]{/*}{*/},
}

\lstdefinelanguage{smt}{
  sensitive = false,
  keywords={declare, fun, synth, check, assert, sat, blocking, define, constraint},
  numbers=left,
  numberstyle=\footnotesize,
  stepnumber=1,
  numbersep=8pt,
  showstringspaces=false,
  breaklines=true,
  frame=top,
  comment=[l]{;},
}

\definecolor{codegreen}{rgb}{0,0.6,0}
\definecolor{codegray}{rgb}{0.5,0.5,0.5}
\definecolor{codepurple}{rgb}{0.58,0,0.82}
\definecolor{backcolour}{rgb}{1.0,1.0,1.0}

\lstdefinestyle{uclidstyle}{
  backgroundcolor=\color{backcolour},
  commentstyle=\color{codegreen},
  keywordstyle=\color{magenta},
  numberstyle=\tiny\color{codegray},
  stringstyle=\color{codepurple},
  basicstyle=\footnotesize,
  breakatwhitespace=false,
  basicstyle=\footnotesize\ttfamily,
  breaklines=true,
  captionpos=b,
  keepspaces=true,
  numbers=left,
  numbersep=5pt,
  showspaces=false,
  showstringspaces=false,
  showtabs=false,
  tabsize=2,
  frame=shadowbox
}

\begin{document}
\title{Synthesis in \uclid}

\author{Federico Mora}
\affiliation{
  \institution{University of California, Berkeley}
}
\author{Kevin Cheang}
\affiliation{
  \institution{University of California, Berkeley}
}
\author{Elizabeth Polgreen}
\affiliation{
  \institution{University of California, Berkeley}
}
\author{Sanjit A. Seshia}
\affiliation{
  \institution{University of California, Berkeley}
}

\begin{abstract}
    We describe an integration of program synthesis into \uclid, a formal modelling and verification tool. To the best of our knowledge, the new version of \uclid is the only tool that supports program synthesis with bounded model checking, k-induction, sequential program verification, and hyperproperty verification. We use the integration to generate 25 program synthesis benchmarks with simple, known solutions that are out of reach of current synthesis engines, and we release the benchmarks to the community. 
\end{abstract}

\maketitle

\section{Introduction}\label{sec:intro}
Formal verification can be a time-consuming task that requires significant manual effort.
Especially for complex systems, users often need to manually provide, for example, loop invariants, function summaries, or environment models. 
Synthesis has the potential to alleviate some of this manual burden~\cite{seshia-pieee15}. 
For example, prior work has used synthesis to reason about program loops~\cite{DBLP:conf/lpar/DavidKL15}, and to automate program repair~\cite{DBLP:conf/sigsoft/LeCLGV17}.
We believe this is a promising direction, but, for it to make a real impact,
verification tools need to offer flexible synthesis integration, generic support for proof procedures, and a capable synthesis engine back-end.

In this work, we primarily address the first two requirements. Specifically, we integrate program synthesis into the \uclid~\cite{uclid5} formal modelling and verification tool by allowing users to declare functions to synthesize and to use these functions freely. While \uclid has previously
supported use of synthesis, it only supported invariant synthesis through a special command 
that was independent of verification.
We use the new synthesis integration to generate 25 benchmarks from existing verification tasks. These benchmarks have small solutions, but are out of reach for current synthesis engines. 
We hope that they will help the synthesis engine development effort, particularly for syntax-guided synthesis~\cite{alur-fmcad13}.

\paragraph{Illustrative Example}
Consider the \uclid model in Fig.~\ref{ex:fib}, which represents a Fibonacci sequence. 
The (hypothetical) user wants to prove by induction that the invariant \codelike{a\_le\_b} at line 13 always holds. Unfortunately, the proof fails because the invariant is not inductive. Without synthesis, the user would need to manually strengthen the invariant until it became inductive. However, the user can ask \uclid to automatically do this for them. Fig.~\ref{ex:fib} demonstrates this on lines 16, 17 and 18. Specifically, the user specifies a function to synthesize called \codelike{h} at lines 16 and 17, and then uses \codelike{h} at line 18 to strengthen the existing set of invariants. Given this input, \uclid, using e.g. \textsc{cvc4}~\cite{cvc4} as a synthesis engine, will automatically generate the function \codelike{h(x, y) = x >= 0}, which completes the inductive proof. 

In this example, the function to synthesize represents an inductive invariant. However, functions to synthesize are treated exactly like any interpreted function in \uclid: the user could have called \codelike{h} anywhere in the code. Furthermore, this example uses induction and a global invariant, however, the user could also have used a linear temporal logic (LTL) specification and bounded model checking (BMC). In this sense, our integration is fully flexible and generic.

\paragraph{Contributions} 
We present an integration of synthesis into the verification tool \uclid, allowing users to generate program synthesis queries for unknown parts of a system they wish to verify. The integration is a natural extension of the existing \uclid language, and to the best of our knowledge, is the first to support program synthesis with bounded model checking, k-induction, sequential program verification, and hyperproperty verification. The synthesis queries \uclid generates are in the standard \sygusif~\cite{sygus-if} specification language. We use this tool to generate a 25 \sygusif synthesis benchmarks from existing verification queries and release these benchmarks to the community. 

\section{Related work}
Program sketching~\cite{solar2009} synthesizes expressions to fill holes in programs, and has subsequently been applied to program repair~\cite{DBLP:conf/sigsoft/LeCLGV17,DBLP:conf/icse/HuaZWK18}. \uclid aims to be more flexible than this work, allowing users to declare unknown functions even in the verification annotations, as well as supporting multiple verification algorithms and types of properties.
Rosette~\cite{DBLP:conf/oopsla/TorlakB13} provides support for synthesis and verification, but the synthesis is limited
to bounded specifications of sequential programs, whereas \uclid can also synthesize programs that satisfy unbounded specifications, by using proof procedures like induction.
Formal synthesis algorithms have been used to assist in verification tasks, such as safety and termination of loops~\cite{DBLP:conf/lpar/DavidKL15}, and generating invariants~\cite{DBLP:conf/tacas/FedyukovichB18,DBLP:conf/vmcai/ZhangYFGM20}, but none
of this work to-date integrates program synthesis fully into an existing verification tool. 
Before this new synthesis integration, \uclid supported synthesis of inductive invariants. The key insight of this work is to generalize the synthesis support, and to unify all synthesis tasks in \uclid by re-using the verification back-end.

\section{From Verification to Program Synthesis}
In this section, we give the necessary background on program synthesis, and the existing verification techniques inside of \uclid. We then describe how we combine the two to realize synthesis in \uclid. 

\begin{figure}[t]
\begin{lstlisting}[language=uclid,style=uclidstyle]
module main {
  // Part 1: System Description.
  var a, b : integer;
  init {
    a, b = 0, 1;
  }
  next {
    a', b' = b, a + b;
  }

  // Part 2: System Specification.
  invariant a_le_b: a <= b; 
  
  // Part 3: (NEW) Synthesis Integration
  synthesis function 
    h(x : integer, y : integer): boolean;
  invariant hole: h(a, b);

  // Part 4: Proof Script.
  control {
    induction;
    check;
    print_results;
  }
}
\end{lstlisting}
  \caption{\uclid Fibonacci model. Part 3 shows the new synthesis syntax, and how to find an auxiliary invariant.}
  \label{ex:fib}
\end{figure}
  
\subsection{Program Synthesis}
The program synthesis problem corresponds to the second-order query $$\exists f\; \forall \vec{x} \sigma(f, \vec{x}) \label{eq:sygus},$$ where $f$ is the function to synthesize, $\vec{x}$ is the set of all possible inputs, and $\sigma$ is the specification to be satisfied.

\subsection{Verification in \uclid}
At at high level, \uclid takes in a model, generates a set of verification conditions, asks a satisfiability modulo theory (SMT) solver~\cite{barrett-smtbookch09} to check the verification conditions, and then returns the results to the user. This process is the same regardless of the proof procedure used. The important point, is that \uclid encodes the violation of each independent verification condition as a separate \smtlib query.

Let $P_i(\vec{x})$ encode the $i^{th}$ verification condition, and take the $i^{th}$ \smtlib query to be checking the validity of $\exists \vec{x} \; \neg P_i(\vec{x})$, where $P_i$ contains no free variables. 
We say that there is a counterexample to the $i^{th}$ verification query if the query $\exists \vec{x} \; \neg P_i(\vec{x})$ is valid. Verification of a model with $n$ verification conditions succeeds \emph{iff} there are no counter-examples:
$$\forall \vec{x}\; \bigwedge_{i=0}^{i=n} P_i(\vec{x}).$$

\subsection{Synthesis Encoding in \uclid}\label{sec:encoding}
Given a \uclid model in which the user has declared a function to synthesize, $f$, we wish to construct a synthesis query that is satisfied \emph{iff}
there is an $f$ for which all the verification conditions pass for all possible inputs. We build this synthesis query by taking the conjunction of the negation of all the verification queries. Specifically, we check the validity of
$$\exists f \; \forall \vec{x}\; \bigwedge_{i=0}^{i=n} P_i(f, \vec{x})$$
where each $P_i$ encodes a verification condition that may refer to the function to synthesize, $f$. Note the similarity between this query and the standard program synthesis formulation: the specification for synthesis, $\sigma$, from the equation in Sec.~\ref{eq:sygus}, is now replaced with the conjunction of all the verification conditions. With this observation, to enable synthesis for any verification procedure in \uclid, all we do is let users declare and use functions to synthesize.

\section{Implementation}
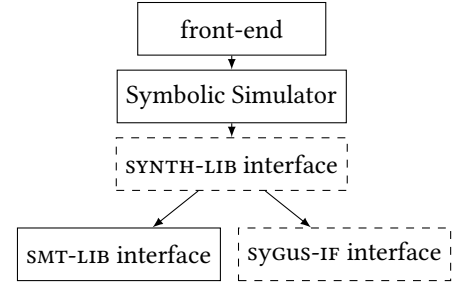
\begin{figure}[t]
\begin{tikzpicture}[>=latex,x=3cm,y=2cm]
\node[rectangle,draw,minimum height=0.7cm,minimum width=2.5cm,align=center] at (1,1) (f) {front-end};
\node[rectangle,draw,minimum height=0.7cm,minimum width=2.5cm,align=center] at (1,0.55) (s) {Symbolic Simulator};
\node[rectangle,draw,dashed, minimum height=0.7cm,minimum width=2.5cm,align=center] at (1,0.1) (l) {\llamalib interface};
\node[rectangle,draw,minimum height=0.7cm,minimum width=2.5cm,align=center] at (0.5,-0.5) (smt) {\smtlib interface};
\node[rectangle,draw,dashed,minimum height=0.7cm,minimum width=2.5cm,align=center] at (1.5,-0.5) (sygus) {\sygusif interface};
\path[->] (f) edge node {} (s);
\path[->] (s) edge node {} (l);
\path[->] (l) edge node {} (smt);
\path[->] (l) edge node {} (sygus);
\end{tikzpicture}
\caption{Overview of synthesis in \uclid. Dashed blocks indicate blocks introduced for the new synthesis integration.}
\label{fig:uclid}
\end{figure}

The \uclid verification tool is constructed as shown in Figure~\ref{fig:uclid}. An input \uclid model is parsed by the front-end into an abstract syntax tree.
From this abstract syntax tree, a symbolic simulator generates an assertion stack that contains an assertion for each verification condition. 
Prior to our work, assertions were then passed to an \smtlib interface which converted the assertions to \smtlib and called a solver. The new \uclid instead uses a new intermediate representation, \llamalib, that is easily passed to either an SMT solver or a synthesis engine. This architecture allows us to use the same code that generates verification queries for synthesis.

The \llamalib representation is \smtlib \cite{smt2}, but with one extra command borrowed from \sygusif \cite{sygus-if}. The syntax for the new command is
\begin{align*}
(\text{\keyword{synth-blocking-fun}}& \; \langle \text{fname} \rangle \; \\((
\langle \text{argname} \rangle \; &\langle \text{argsort} \rangle )^* ) \;
\langle \text{rsort} \rangle \; \langle \text{grammar}\rangle?),
\end{align*}
where $\langle \text{fname} \rangle$ is the name of the function, $\langle \text{argname} \rangle$ is the name of an argument, $\langle \text{argsort} \rangle$ is the sort of the corresponding argument, there are zero or more arguments, $\langle \text{rsort} \rangle$ is the sort returned by the function, and $\langle \text{grammar} \rangle$ is an optional syntactic specification for the function body.
Intuitively, a \llamalib query with a single \keyword{synth-blocking} \keyword{-fun} declaration asks ``is there a function that makes this underlying \smtlib query unsatisfiable?''

Fig.~\ref{ex:fib-synth} shows the \llamalib query corresponding to the Fibonacci model in Fig.~\ref{ex:fib}. A synthesis engine might solve the query in Fig.~\ref{ex:fib-synth} by finding the function \codelike{h(x, y) = x >= 0}. This is a correct solution because the corresponding \smtlib query---which we can get by commenting out line 1 of Fig.~\ref{ex:fib-synth} and uncommenting line 2---is unsatisfiable.

\begin{figure}[t]
\begin{lstlisting}[language=smt,style=uclidstyle]
(synth-blocking-fun h ((x Int) (y Int)) Bool)
;(define-fun h ((x Int) (y Int)) Bool (>= x 0))
(declare-fun initial_b () Int)
(declare-fun initial_a () Int)
(declare-fun new_a () Int)
(declare-fun new_b () Int)
(assert (or 
 (not (and (<= initial_a initial_b) (h 0 1)))
 (and 
  (and (<= initial_a initial_b) (h initial_a initial_b))
  (= new_a initial_b)
  (= new_b (+ initial_a initial_b))
  (not (and (<= new_a new_b) (h new_a new_b))))))
(check-sat)
\end{lstlisting}
    \caption{\llamalib induction query of Fig.~\ref{ex:fib}}
    \label{ex:fib-synth}
\end{figure}

The semantics of \llamalib is exactly that of \smtlib when no function to synthesize is on the assertion stack, and assertions are passed directly to the SMT solver. 
When the assertion stack contains a function to synthesize, \uclid applies the following four rewrite rules
to convert \llamalib into \sygusif:
\begin{enumerate}[leftmargin=*]
    \item (assert $a$) $\rightarrow$ (constraint (not $a$))
    \item (declare-fun $a$ ($s_0 ... s_{n-1}$) $s_n$ $\rightarrow$ (declare-var $a$ $s_0$ ->...-> $s_n$)
    \item synth-blocking-fun $\rightarrow$ synth-fun
    \item check-sat $\rightarrow$ check-synth
\end{enumerate} 
The first rewrite rule is the most important: it implements the following equivalence 
$$\exists f \; \neg \exists \vec{x}\; \bigvee_{i=0}^{i=n} \neg P_i(f, \vec{x}) \equiv
\exists f \; \forall \vec{x}\; \bigwedge_{i=0}^{i=n} P_i(f, \vec{x}),$$
where the left hand side is the form of queries in \llamalib, and the right hand side is the corresponding query in \sygusif. The source code for \uclid is available online \cite{source}.


\section{Benchmark Suite}\label{sec:results}
The integration of synthesis into \uclid allows us to generate synthesis benchmarks from any \uclid verification task. We thus present a set of 25 benchmarks with known, small solutions that are out of reach of existing synthesis solvers. These benchmarks use induction, BMC, LTL specifications, and sequential code. To conform to the \sygusif language, we limited ourselves to bit-vector, integer, array, and boolean data-types, and did not use verification tasks that required quantifiers. All benchmarks are available online \cite{bench}.

The benchmarks come from four different sources. Four benchmarks come from a simplified model of the Two Phase Commit protocol, written in P~\cite{desai2013p}; three benchmarks come from Sahai at al's~\cite{sahai2020verification} work on hyperproperty verification; six benchmarks come from \uclid's documentation; and the remaining 12 benchmarks come from models used in UC Berkeley's EECS 219C course. In all cases, we constructed the benchmarks by replacing small parts of either auxiliary invariants or parts of existing code with functions to synthesize. 12 benchmarks come from models that use induction, and 13 from models that use LTL specifications and BMC. All 25 benchmarks are difficult for existing state-of-the-art engines, but are a reasonable target for synthesis engines.

\section{Conclusions and Future Work}\label{sec:futurework}
We have presented an integration of synthesis into the \uclid verification tool, allowing users to generate synthesis
queries for unknown parts of a system they wish to verify. This integration is compatible with all verification algorithms currently supported by \uclid, and generates synthesis queries in the standard \sygusif format.

In the future, we intend to apply synthesis in \uclid to the verification of distributed systems written in P. Prior work has been successfully in finding invariants for bounded distributed systems, and then generalizing the invariants to the unbounded setting~\cite{I4}. We plan to explore these approaches with \uclid now that we can easily switch between synthesis using e.g. BMC and k-induction.

\begin{acks}
This work was supported in part by NSF grants 1739816 and 1837132, a gift from Intel under the SCAP program, SRC Task 2867.001, and the iCyPhy center.
\end{acks}

\clearpage
\balance

\bibliographystyle{ACM-Reference-Format}
\bibliography{paper}

\end{document}